\documentclass{article}

\usepackage[nonatbib]{arxiv}

\usepackage[utf8]{inputenc} 
\usepackage[T1]{fontenc}    
\usepackage{hyperref}       
\usepackage{url}            
\usepackage{booktabs}       
\usepackage{amsfonts}       
\usepackage{nicefrac}       
\usepackage{microtype}      
\usepackage{lipsum}
\usepackage{listings}  
\usepackage{graphicx}
\usepackage[calcwidth = .9\linewidth, font = small, labelfont = bf]{caption}

\title{Promoting Distributed Trust in Machine Learning and Computational Simulation via a Blockchain Network}

\makeatletter
\def\ScaleIfNeeded{%
\ifdim\Gin@nat@width>\linewidth
\linewidth
\else
\Gin@nat@width
\fi
}
\makeatother

\author{
  Nelson Kibichii Bore,$^1$ Ravi Kiran Raman,$^{1,2}$ Isaac M. Markus,$^1$ Sekou L. Remy,$^1$ \\\bf Oliver Bent,$^1$ Michael Hind,$^1$ Eleftheria K. Pissadaki,$^1$ Biplav Srivastava,$^1$\\ \bf Roman Vaculin,$^1$ Kush R. Varshney,$^1$ and Komminist  Weldemariam$^1$  \\
 $^1$IBM Research \\
       $^2$University of Illinois at Urbana-Champaign \\
}

\begin{document}
\maketitle

\begin{abstract}
Policy decisions are increasingly dependent on the outcomes of simulations and/or machine learning models. The ability to share and interact with these outcomes is relevant across multiple fields and is especially critical in the disease modeling community where models are often only accessible and workable to the researchers that generate them. This work presents a blockchain-enabled system that establishes a decentralized trust between parties involved in a modeling process. Utilizing the OpenMalaria framework, we demonstrate the ability to store, share and maintain auditable logs and records of each step in the simulation process, showing how to validate results generated by computing workers.  We also show how the system monitors worker outputs to rank and identify faulty workers via comparison to nearest neighbors or historical reward spaces as a means of ensuring model quality.

\end{abstract}


\section{Introduction} \label{sec:introduction}

Technological innovation has facilitated large-scale computation where  data collection, management, analysis, and interpretation can be achieved as a multi-party process. The complexity of having multiple organizations or participants involved in the collection, utilization and modeling of data, including policies derived from such, results in an increased risk of under-resourced, negligent, or bad faith parties compromising the entire data and modeling pipeline. One domain where this scenario is highly relevant is disease modeling, where simulations are often performed in a distributed fashion, with both inputs and outputs generated or consumed by remote parties in under-resourced settings.

Let us consider a scenario where a disease modeling scientist (DMS) is interested in running experiments to simulate malaria epidemiology and control using a computational modeling agent, e.g., OpenMalaria (OM) \cite{18694530}. To study the infection dynamics for a geographic region, the DMS defines parameters describing the scenario, including demographics, entomology, health system and intervention policies.  The DMS will focus on evaluating different interventions such as distributing insecticide treated nets 
and commissioning indoor residual spraying 
in varying proportions. The OM model evaluates the efficacy of these possible intervention policies in terms of disability-adjusted life years 
(total productive life years lost related to malaria infection). In addition, the policies are also evaluated based on the associated cost of implementation, e.g., health system cost 
and intervention costs.  These factors are a great concern in malaria modeling and decision making where any misinterpretation or compromise of an intervention in the model could have far-reaching public health consequences. Given the computational expense of modeling and computations, it is desirable for various decision makers (e.g.,\ government health ministries and non-governmental organizations) to rely on efficacy evaluations computed by others without re-running models themselves, which may be prohibitively expensive.

This scenario points to the need for both algorithmic accountability and decision provenance such that simulation or machine learning systems can,  at a minimum, be audited and tracked, and preferably have results validated before actions or policies are derived \cite{arifin2015verification, ferris2015openmalaria, bent2017novel}.  In defining the requirements of such systems, we consider two distinct functionalities. 

The first functionality is logging each interaction and metadata associated with every step in the data process pipeline, from data collection to model creation, utilization and results sharing. This includes identifying participants or organizations that interact with data and  models, the methodologies and steps used in generating, handling or manipulating and sharing the data and/or models, the resources used in the process, and any other read, write or execute activities. 

The second functionality is a validation scheme which can enable users or systems to accept, reject or require further actions. This functionality is intended to provide multiple layers of enhanced trust regarding data and model provenance. 
The two functionalities described are intended to explicitly or implicitly address several growing concerns around data, models and derived policies, including reproducibility, access, provenance and reliability.

Beyond disease modeling, several computational fields face many of common concerns due to the rapid rate of development and progress. From the pervasiveness of ad-hoc methodologies, to weak or unreliable access controls to data, to ultimately faulty conclusions reached through bias or tainted models. The need to provide added layers of trust is crucial for technical, practical and ethical/legal reasons \cite{varshney2017safety,hind2018trust}. However, the ability to implement the required overhead that addresses these concerns is neither mature nor reliable enough to gain wide adoption among the respective scientific communities. 

There are efforts to answer a few of the above concerns from a system perspective. The thesis \cite{adebayo2016fairml} presents FairML, an end-to-end toolbox for evaluating predictive model fairness by examining their inputs and treating the algorithm as a black box, quantifying some of the cost associated with promoting fairness in a system. Other end-to-end machine learning pipelines have been proposed where there are multiple entities interacting with data, models and policies \cite{Shaikh2017}.  However, these systems mainly assume that (computational) workers executing simulations or learning tasks can be relied upon to deliver trusted results, where the fault would mainly lie at the algorithm or data preparation level. As it is possible to envision a system where remuneration is paid commensurate with computation, it becomes essential to extend validation and verification schemes to the output produced by workers to maintain system integrity. A trusted system should address, at least, the following three questions:  

\begin{itemize}
\item[Q1:] 
How can we facilitate the sharing of data and models securely among different modeling communities where computation has been performed remotely? 

\item[Q2:] Can we detect and eliminate invalid results and workers in the system to establish a digital trust policy among participating entities? 

\item[Q3:] What is the computational cost to achieve the digital trust when checks are occurring at data, algorithm, and worker level?  
\end{itemize}

This work presents a blockchain-enabled distributed system to address the above questions by establishing and promoting a decentralized trust between parties involved in disease modeling.   The prototype system is designed and implemented to facilitate (i) the sharing of simulation and/or machine learning assets securely and (ii) validation of results or outputs of simulation or machine learning computations. 
The initial setup of the blockchain network mimics a disease modeling community consisting of several organizations 
with whom we have extensively engaged during the course of developing the prototype system and specifically based on our active involvement in eradicating malaria in developing countries \cite{bent2017novel}. 
In this blockchain network, each organization can operate its own blockchain node (peer) with its own copy of the shared, distributed and immutable ledger (database) which records all the audit logs as will be explained later. Importantly, potentially every organization running the blockchain peer can  independently validate the process and submitted results as well as participate in identifying invalid or problematic results.

The rest of the paper is structured as follows. Section \ref{sec:background} motivates this work by presenting background materials. Section \ref{sec:systemoverview} discusses the various components and their implementation. Section \ref{sec:experiment} describes the experimental setup and Section \ref{sec:results} presents results of our experiments. Finally, in Section \ref{sec:conclusion}, we conclude the work and also provide possible future directions.

\section{Background and Motivation}
\label{sec:background}

Data and decision provenance are key considerations in the growing literature of algorithmic fairness, accountability,  and transparency (FAT). Beyond the methodologies developed by researchers to ensure the former, there are several considerations around the infrastructure required to promote accountable and transparent models. End-to-end systems have been discussed by \cite{Shaikh2017,BowerKNSVV17}, noting required components and possible architectures for capturing and handling events in the modeling and decision making process. 

Reference \cite{Veale2018} explored a different dimension of model provenance and usage under the recent General Data Protection Regulation (GDPR) legislation, examining how not just data but also model provenance and access needs to be monitored due to the possibility of model inversion and inference attacks. For entities providing machine learning models as a service (e.g., DLaaS \cite{DBLP:journals/ibmrd/BhattacharjeeBD17}, MLaaS \cite{DBLP:conf/icmla/RibeiroGC15}), the liability exposure implies certain infrastructure requirements such as defined storage limitation (keeping data for only as long as needed), erasure rights, and information rights. Furthermore, it can be argued that under GDPR legislation, remote computation needs to be monitored and tracked in case third parties are contracted for analyzing and using data to build models.

Public policy FAT considerations also have an established literature in the field of decision-making or decision-support systems \cite{zhengmeng2011brief, lepri2017fair}. Since the 1970s these types of systems have been introduced into multiple industries from manufacturing to health care  to provide improved visibility and understanding into business process rules. Another more recent movement has been around the development of Open Algorithms under the OPAL project from MIT. This work outlines multiple challenges with providing trusted models, from secure member on-boarding, to data sharing processes, to worker remuneration \cite{DBLP:journals/corr/HardjonoP17, hardjono2016trust, basakfortifid}. Additionally, \cite{Dinh2018} surveyed blockchains as data processing platforms.

Further motivation for this work comes from the critical outcomes associated with the disease modeling community \cite{ferris2015openmalaria, arifin2015verification, 8416392} where sub-optimal outcomes have direct consequences to human lives. Unfortunately, depending on the specific disease modeled, geographical considerations create complicate scenarios for model sharing and development. Due to a lack of computing resources and a fragmented field, researchers are not able to access or further develop models not generated by themselves. Furthermore, the validation and verification (V\&V) process can be impaired, since acquisition of real world data is difficult, costly or unreliable. Lastly, the actual complexity involved in evaluating the reward and policy space for managing disease transmission and/or prevention, requires high performance computing, which augments the needs for precise V\&V schemes to prevent unnecessary expenditure of computing resources. 
Although there are various V\&V methods at the data and model level \cite{niaz2010verification, niaz2010p,arifin2015verification}; this work explores the utilization of worker outputs for doing individual assessment of worker outputs. 

Finally, blockchain technology \cite{Nakamoto_bitcoin:a,Androulaki} provides a decentralized platform with the objective of improving transparency, data integrity and security of transactions among involved participants \cite{Swan:2015:BBN,Tapscott:2016:,Hari:2016:IBD,Androulaki}.    
Blockchain platforms are based on distributed, shared, immutable ledgers with each participant of the blockchain network possibly having its own copy of the data. The algorithms and protocols of blockchain maintain data replication and consistency of each ledger copy, provide support for recording blockchain data and transactions, assure the integrity of transactions by providing mechanisms for achieving consensus among the participants, guarantee transactions finality and assure that transactions are executed by identifiable entities (non-repudiation). Additionally, generic blockchain platforms, such as Hyperledger Fabric \cite{Androulaki}, also support execution of business logic along with the transactions in the form of chaincode (or so called smart contracts) which typically have the form of an executable code in languages such as JavaScript, GoLang, etc. Given the above properties, blockchain platforms and solutions have been particularly suitable for problems in decentralized settings where there is a need for increased visibility, need of compliance or lack of trust. 

In the case of our particular problem, all entries of, and changes to, records on blockchain ledger  represent the entities associated with computational or machine learning assets and all relevant operations applied to these assets. Because of the built-in immutability and consensus mechanisms of blockchain, any alteration of an asset (e.g., trained model or simulation result) can be verified against a particular instance of the record on blockchain. Any participant of the blockchain network can audit the records and their validity at any time. Our work additionally contributes distributed validation and verification mechanisms specifically suitable for computational assets and processes, and it leverages a novel compression schema \cite{RamanVHRPBDSV2018_arxiv} built into the pipeline to address some scalability concerns of blockchains \cite{Meiklejohn2018}.  

\section{The System}
\label{sec:systemoverview}
In this section, we provide a brief overview of the proposed system shown in Figure~\ref{fig:system} and its prototype implementation.

\subsection{Overview}
The system leverages blockchain to allow users to track, validate and verify machine learning task and results. These include models, outputs, policies, model parameters, and other meta-data associated with the learning process. Each of these can be treated as an asset in a blockchain network, defining a relationship schema as a template for process or provenance engine to map the transition of assets across a blockchain network. In addition, the system has a trusted service that the system uses to validate and verify worker outputs. The core components of the system include: External Systems of Engagement, Process Engine, Message Processor, Learning Layer, Notification Manager and Validation Engine.

\begin{figure}
	\centering
	  \includegraphics[width = 0.5\linewidth]{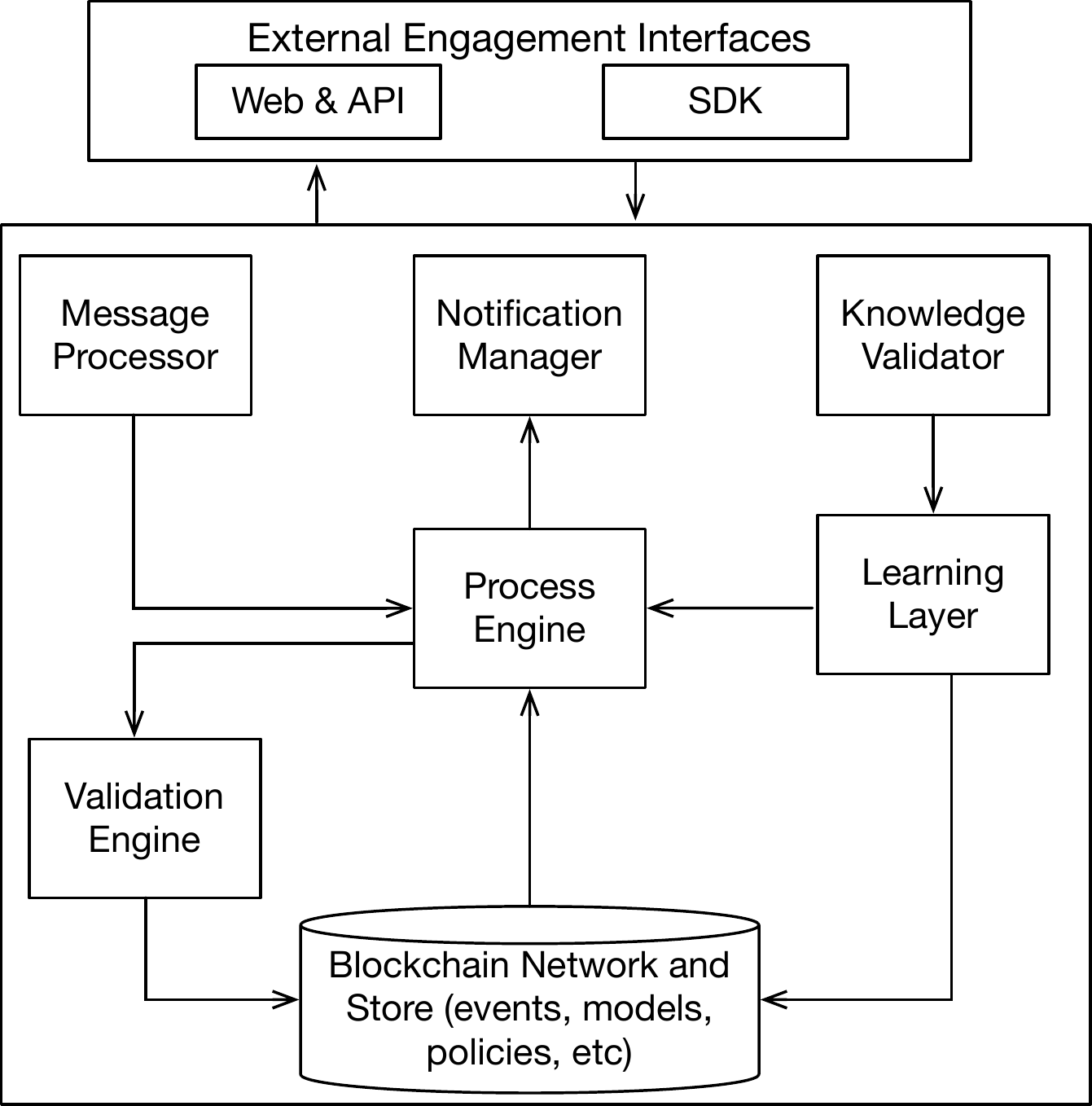}
	\caption{High-level view of the system components}
	\label{fig:system}
\end{figure}
The \textit{external engagement interfaces} exposes a set of cloud-enabled APIs. The APIs serve as input/output end-points to enable users to post their experiments, receive real time updates of policies simulation progress and view validated simulation results. Some modeling or simulation requirements or needs (e.g., due to country or organization-specific privacy acts)  may restrict the sharing of sensitive data and models on cloud-enabled infrastructure. In such cases, the system can be packaged as a software development kit (SDK) and deployed on private on-premise infrastructure. The SDK comes with bundled underlying services as shown in Figure~\ref{fig:system}, such that engineers can configure appropriately for on-premise experiment setup. Some of the configurations that they can perform include configuration of the application with their on premise blockchain network and still leverage the validation engine. 

The \textit{process engine} is responsible for executing model policies based on a specific executor (e.g., OpenMalaria \cite{18694530} or EMOD \cite{ferris2015openmalaria}) that the users specify in their simulation setup. The framework allows a user to upload their executor model, which is then available for other users or organizations who are onboarded onto the blockchain network. The shared executor model is securely stored and managed on the blockchain network to guard from malicious intentions or users through inbuilt cryptography \cite{DBLP:conf/eurosp/HalpinP17}. The process engine receives experiment requests from a message queue (e.g., Kafka or RabbitMQ) and executes the experiment based on the executor model logic. The experimental results are thereafter securely stored on the blockchain network after they have been validated by the validation engine.  Also note that the process engine is a decentralized micro-service configured to each node in our blockchain network. This service  facilitates the end-to-end experience  of model runs as discussed in this paper.

The \textit{message processor} is responsible for receiving tasks from external engagements interfaces. This component filters simulation payloads for duplicates to avoid resource wastage and also determines the degree of similarity from previously posted model policies. In the event that there exist simulation duplicates, this component will inform the process engine to retrieve the previous simulation results for further adjustments if needed. For new simulations this component sends the information to the process engine on how to execute simulations and which existing policy results to learn and validate from during model execution. This information is also stored securely on the blockchain for validity and authenticity of the data.

The \textit{blockchain layer} is responsible for ensuring decentralized trust among the parties involved in model creation, execution, simulation and validation. This includes policies, simulation results, interaction events, validation data and executors being stored in a secure way to ensure that they are valid, secure and authentic. The component is also used in managing users and creating a long-lasting trackable reference that others will be able to consult when their model policies match. The use of peers in a Hyperledger Fabric (HLF) \cite{Androulaki} blockchain network also ensures that the policies and simulation results are validated by human input to make them effective and useful.

We leverage specialized smart contracts/chaincodes in HLF (i.e., system chaincodes) that manage system level business logic and parameters as used  during consensus, endorsement, etc.  These chaincodes are trusted distributed applications which leverage tamper-proof properties of the blockchain and an underlying agreement between blockchain nodes which is referred to as an endorsement or endorsement policy \cite{Androulaki}. In general, blockchain transactions  must be “endorsed” before being committed to the blockchain while transactions which are not endorsed are disregarded. A typical endorsement policy allows chaincode to specify endorsers for a transaction in the form of a set of peer nodes that are necessary for endorsement. When a client sends the transaction to the peers specified in the endorsement policy, the transaction is executed to validate the transaction. After validation, the transactions enter an ordering phase in which a consensus protocol is used to produce an ordered sequence of endorsed transactions grouped into blocks. 

The \textit{learning layer} is responsible for providing a service where different models can leverage other results from other models that are deemed to be  similar based on their inputs and target model. This depends on the asset relationship schema stored on the blockchain ledger to externalize any events and metadata for other model simulation activities in an attempt to use trust data and observations while reducing unnecessary computing cost. 
The usefulness of the existing simulations output is determined by users that give their report via knowledge validator to the learning layer. 

The \textit{notification manager component} coordinates the communication between users who have posted their model policies for execution and informing them once completed. This component is also useful when the learning layer requires user validation on new model policies for learning.

The \textit{validation engine} is responsible for ensuring consistency and validity of experiment simulation results. Experimental simulations and computations for machine learning tasks produce a large volume of data that are complex to validate manually. Adopting the notion of valid computing from \cite{RamanVHRPBDSV2018_arxiv}, a reported computation is declared valid if it is close to the recomputed estimates. Drawing insight from the distributed endorsements used by cryptocurrencies, the validation engine, upon receiving computation results from the process engine, selects a set of peers, called endorsers, who recompute the report and endorse the computation if the deviation from recomputation is within an acceptable tolerance. Considering the challenge of system scalability owing to the communication and computation overheads, the validation engine employs the compression schema of \cite{RamanVHRPBDSV2018_arxiv} for approximate reporting and validation. Given secure access to the computational models and parameters, the endorsers can perform parallel validation through recomputation, while accounting for any randomness in the computational pipeline. The validated states, upon achieving consensus on the endorsements are then added to the blockchain ledger to maintain an audit of validated computations. This record of validated computations allows for simple verification by other agents who just need to ensure consistency of the ledger (hash chain) to verify the results. Appropriate endorsement consensus policies and validation tolerance can be encoded into the validation engine depending on the extent of trust and accuracy desired in the computation.

\subsection{Implementation}
\label{sec:implementation}

\begin{figure}[h!]
	\centering
	  \includegraphics[width = \linewidth]{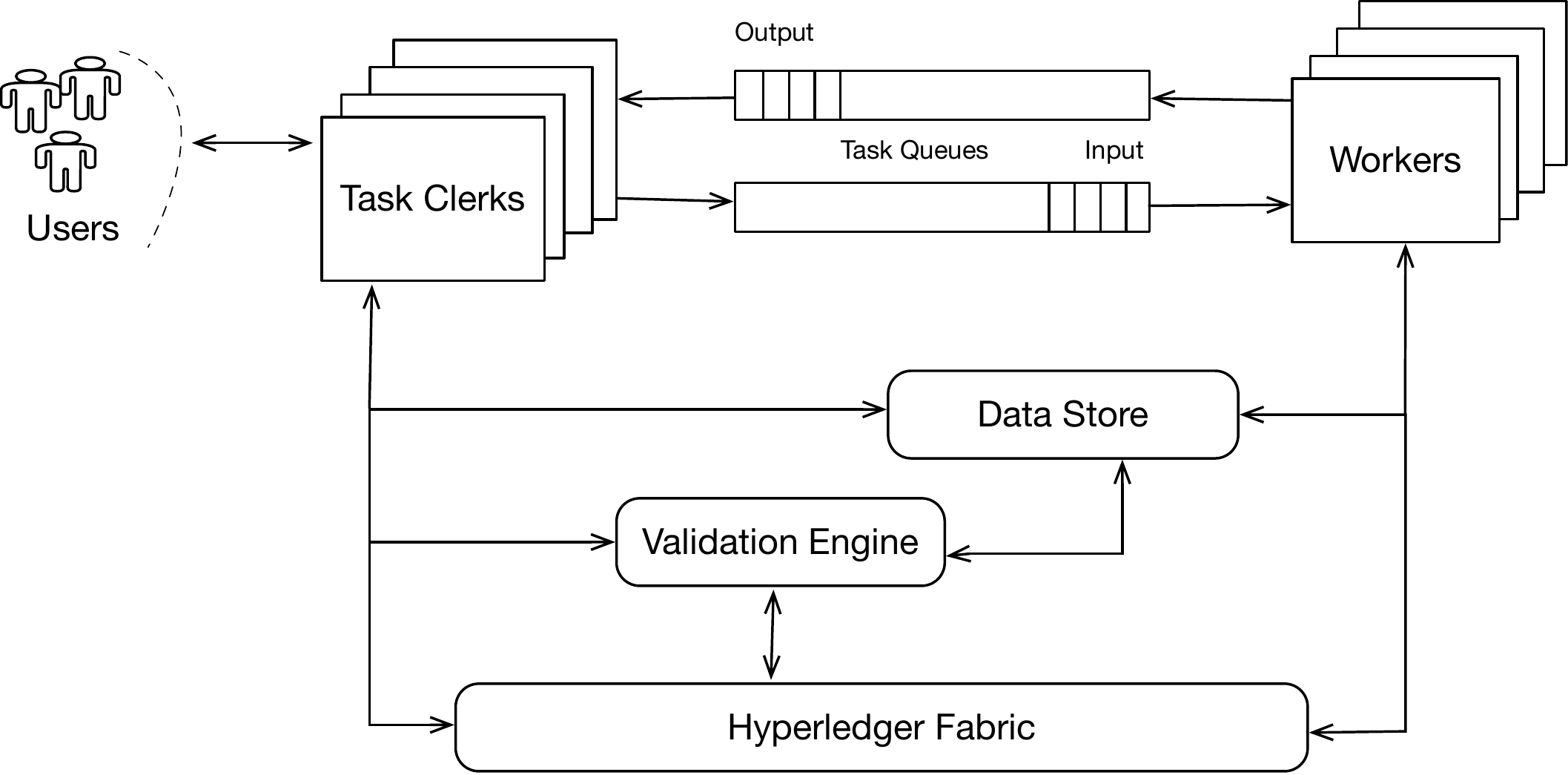}
    \caption{An example of a deployment instance}
    \label{fig:implementation}
\end{figure}

We implemented the system as a suite of microservices running on docker containers (see Figure \ref{fig:implementation}) to enable running on different computing environments. Each computational worker is also packaged as a docker container containing  OM executable, application and scripts to engage with message scheduler using RabbitMQ to consume input files, and post the output files as needed to task clerk. During deployment, the workers are capable of executing different models compatible with the OM framework, which can be passed along with the input files required to run a model. The computation results are first stored in the data store for further validation by the validation engine and thereafter persisted to the ledger. In addition to the results, the container also keeps track of the task progress and logs the events to the ledger. Together, these tools are referred to as a worker, and multiple workers can be deployed on a single machine or even distributed across the Internet.

The task clerk exposes an interface where model creators can submit their models to be evaluated by the community by running simulations.  Model evaluators define a specific instantiation of a model that they would like to evaluate in the experiment. The definition will contain the right input specification that aligns with the model of their choice. The resulting file is then stored in the data store with the associated events persisted to the ledger. Thereafter, the task is posted to the task queues ready for computations.

The task request file, as well as the results for the processed task are stored in a scalable distributed non-relational storage. We used IBM's Cloudant \cite{cloudant} to achieve this goal because of its unique nature of handling heavy workloads.  The task clerk, validation engine and workers are connected to the data store for ease of accounting and providing real-time progress of the tasks.

To tie task clerks and workers together, we used a messaging fabric. The current implementation harness Advanced Message Queuing Protocol (AMQP) \cite{Vinoski:2006:AMQ:1187622.1187698} as implemented in the RabbitMQ message queue.  This ensures that the framework can keep track of the tasks completed by every worker. The task clerk posts jobs to a message queue which are subsequently picked up by the idle workers. When the worker completes processing the task, the results are then posted to a different channel on the same queue. This instantiation permits workers to be deployed in a wide range of environments, with little requirements on coordination.

Moreover, the implementatiojn of the blockchain layer is based on the HLF \cite{Androulaki}.  We developed a set of chaincodes to track, store and/or manage data, models and associated metadata as well as the interactions with the assets via process engine. In this work, the blockchain network includes a first, a second, and a third blockchain node. The first blockchain node is configured to receive executor models from one or more users associated with a modeling and create a blockchain transaction including the models’ metadata. The blockchain transaction is configured to store the models, simulation results and metadata to a shared ledger of the blockchain network. The second blockchain node is configured to request for model simulation results validation via validation engine 
based on the executor models. The third blockchain node is configured to compute the model simulation validation based on the executor models, outputs, metadata and stored parameters and provide the validation result to the second blockchain node. The model output validation is based on the validation engine component description in Section  \ref{sec:implementation}. Please note that the current configuration of the blockchain network (i.e., the number of peers/nodes) is only for the purpose of this experiment.

\section{Experiments}
\label{sec:experiment}

To evaluate the proposed system,
consider a scenario where a malaria data scientist (MDS) wishes to  
run simulations on OM to study malaria epidemiology, understand the disease spread and identify optimal disease interventions strategies. The MDS will therefore post simulations in an environment space where there could be a number of anonymous workers. For the purpose of this experiment, we deployed 144 docker containers that were bundled with OM framework as described in the previous section.

An example of a scenario model in OpenMalaria framework is shown in Listing \ref{omscenario}. Users can alter the {\it Interventions} section and run their simulations to determine the efficacy upon implementing the intervention. The simulations posted vary on the execution time which largely depend on the intervention chosen as they may require several iterations to evaluate \cite{ferris2015openmalaria}. Upon execution of the simulation, a \emph{reward} in terms of the cost-adjusted disease affected life years (DALY) is returned by the OM framework. After running the simulations for different intervention policies, the MDS will have to validate and verify these results to look for accuracy and precision which often takes time. Validation and verification can be achieved through different ways e.g. comparing results from other models and use of domain experts \cite{ferris2015openmalaria}. Both approaches are not easily achievable because of the extra resources required for either case (amongst other reasons as discussed in Section \ref{sec:background}). 

\clearpage
\begin{lstlisting}[caption={An example of an OpenMalaria scenario specification},label=omscenario]
{
  <Model (Parametrized with datafrom field studies)>, 
  <Demography (demographics)>,
  <Entomology (specific details including seasonality)>,
  <Health System (case management, treatment seeking, coverage)>,
  <Interventions (e.g., vaccines, Insecticide-Treated bed nets (ITNs), 
  indoor spraying (IRS) or changing health system) >, 
  <...>
}
\end{lstlisting}

To address the challenges of verification and validation, we use our validation engine and in the process try to detect anomalous agents, as evaluated by our experiment here. In our experimental setup to detect anomalies, we model anomalous outputs as noisy rewards, where an input dependent/independent noise is added to the simulated reward. In particular an additive white Gaussian noise, $N \stackrel{i.i.d.}{\sim} \mathcal{N} (c\sigma (ITN - IRS),\sigma)$ where $c$ is a scaling constant and $\sigma$ is the standard deviation of the computations, was introduced to $10\%$ of the workers. The noise represents workers who are biased toward promoting the use of insecticide-treated nets (ITNs) over indoor residual spraying (IRS), and the extent of this bias is characterized by the scaling constant $c$.

The objective of finding anomalous results is to facilitate the verification and validation, and to remove such workers from the system, so as to enable reliable computing. This is crucial in situations where users need to validate large datasets in comparison with other datasets. In addition, it provides a platform to prove that the results and workers may be trusted.

\ifx 
XYZ: trust, anomaly, blockchain

Non-human, trust, anomaly, blockchain, models

Detecting anomalous models in a blockchain network 

Introducing non-human trust in a blockchain network

Identifying trust-able model results/data in a blockchain network

A 
A scalable
Piece of the cost

Words jotted down during Thursday's Biplav meeting

>Person A: I did {X} experiments and the result is f({X})

>Person B: Why should --or how can--  I trust either of those two statements?

>Person B: How can you prove any of this to me?
>Person B: How can I prove any of this myself?

Especially in the setting that the results were generated as a result of Y hours of data collection and Z hours of data processing by some method
>Person C: What are the ways that either statement can be wrong?
>Person C: How can I inject errors in a principled manner to explore how robust a system can be to errors?
>Person C: How much overhead does the system need to provide Type P level of validation/endorsement?
>Person C: What happens if there are also errors in the validation/enforcement processes?
`Given an environment space` How many samples are needed to get an accurate picture of the space? How is that number impacted by the presence of noise? (edited)

for @nelsonbore
\fi

\section{Results}
\label{sec:results}

The experiment was designed with 144 workers running OpenMalaria models distributed across different clusters on IBM Cloud. Each  worker performs a total of $8$ OpenMalaria simulations generating a total of $1152$ valid OpenMalaria results. Of these sources, a randomly sampled subset of $10\%$ of the sources are anomalous and thus generate noisy rewards as described in the previous section. That is, when an anomalous source is sampled, the reward that is returned is the true reward plus a noise $N \sim \mathcal{N} (c\sigma(ITN-IRS),\sigma)$.

In the experiment a client samples $500$ policies, $(ITN, IRS) \in [0,1]^2$ and queries the system for the corresponding reward. The system employs a worker to evaluate the reward corresponding to the policy and the output is validated using the multi-agent blockchain system. Over the course of the experiment, for each source $j \in [1152]$ we keep track of the validation statistic $V_j$ and the deviation record $\Delta_j$ as follows. Let us assume that the output generated by a source $i$ is being validated by an endorser set $\mathcal{E}_i$. Then, for each $j \in \mathcal{E}_i \cup \{i\}$, if the output is valid, update $V_j \leftarrow V_j -1$ and if it is  invalid, $V_j \leftarrow V_j + 1$. Also, for each $j \in \mathcal{E}_i \cup \{i\}$, record the deviation, $\Delta_j \leftarrow \Delta_j \cup \{\delta\}$, where
\begin{equation}
\delta = \left\| \tilde{Y}_i - \frac{1}{m} \sum_{j\in\mathcal{E}_i} \hat{Y}_j \right\|,
\end{equation}
is the endorsement deviation.

We run multiple batches of this experiment and collect the average validation statistics and deviation records for the sources across these batches. We then estimate the probability mass function of the average validation statistic $P_{V_j}$ and that of the deviation $P_{D_j}$ for each source $j$. We use these distributions to detect anomalous sources.

\begin{figure}[h!]
	\centering
	  \includegraphics[width = 0.7\linewidth]{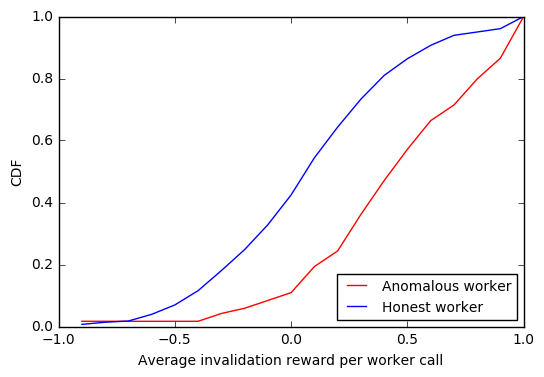}
      \caption{Average invalidation reward per worker call}
\label{fig:advinvalid}
\end{figure}

\begin{figure}[h!]
	\centering
	  \includegraphics[width = 0.7\linewidth]{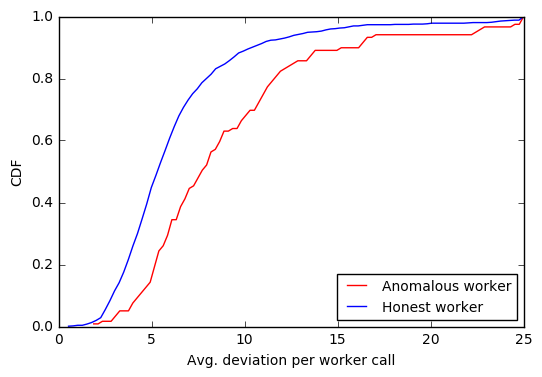}
      \caption{Average deviation per worker call}
          \label{fig:advdevprf}
\end{figure}

Note that all workers involved in a validation cycle record the same validation statistic and deviation. That is, if an anomalous worker is part of the validation cycle, resulting in the invalidation of an output, then all sources involved in that validation incur the corresponding deviation and invalidation. However, as there are only $10\%$ anomalous sources, there is a characteristic difference in the deviation and validation distributions of honest and anomalous workers. This is observed in the form of a separation of the corresponding CDFs, as shown in Figures \ref{fig:advinvalid} and \ref{fig:advdevprf}. 

First, from Figure \ref{fig:advinvalid}, we see that the CDF of anomalous workers is biased in comparison to that of the honest workers. This indicates that anomalous sources incur more invalidations than honest workers as expected as the probability of invalidation in the presence of anomalous sources is higher. We can also observe that the presence of anomalous sources affects the validation statistic of honest workers adversely as well, as they undergo more invalidations than they would otherwise.

Next, we consider the CDFs corresponding to the deviations incurred by the sources in Figure~\ref{fig:advdevprf}. We can again see a clear difference in the distributions of anomalous and honest sources. Also, the anomalous sources incur larger deviations more often than the honest sources as expected. 

We invoke the Kolmogorov-Smirnov (KS) test to check for distributional similarity for two main types of anomalies: biased sources $(c>0)$ and unbiased, input-independent anomalies $(c=0)$. The KS test computes the maximum separation in the CDFs of any two data sources. This statistic is used to compute the p-value corresponding to the binary hypothesis test with null hypothesis that the two sources are identical and the alternate that they are statistically different sources.

\begin{table}[t]
\caption{Kolmogorov-Smirnov test statistics between honest and anomalous workers.} 
\centering
\begin{tabular}{ |c|c|c| } 
 \hline
 {} & Unbiased, $c=0$ & Biased, $c=10$ \\ 
 \hline
 \hline
 {KS stat for $P_V$} & $0.11$ & $0.344$ \\ 
 {p-value for $P_V$} & $0.13$ & $1.9*10^{-11}$ \\ 
 \hline
 {KS stat for $P_D$} & $0.108$ & $0.407$ \\ 
 {p-value for $P_D$} & $0.165$ & $3.4*10^{-16}$ \\ 
 \hline
\end{tabular}
\label{tab:KS_test}
\end{table}

The results of KS test statistic between honest and anomalous workers for both validation and deviation data are shown in Table~\ref{tab:KS_test}. We can observe here that in the biased case, both the validation and deviation profiles are starkly different for honest and anomalous workers, as highlighted by the extremely low p-value. On the other hand, for the unbiased anomalies case, we note that the p-values are comparatively much higher, indicating that whilst different, the statistical confidence in separating the two sources is much lower. This is understandable as detection of anomalies with bias is much easier than unbiased anomalies who just behave like noisy sources.

Next, we consider the task of detection of anomalies using $\{(P_{V_j},P_{D_j}):j \in [1152]\}$ as the feature representations. Treating the problem of anomaly detection as clustering into two clusters, we used the k-NN classifier to determine honest and anomalous workers. In particular, we used the total variational distances between the probability distributions as the notion of distance in the k-NN classifier. Other notions of divergence between probability measures can easily be used without loss of generality.

\begin{figure}
	\centering
	  \includegraphics[width = 0.7\linewidth]{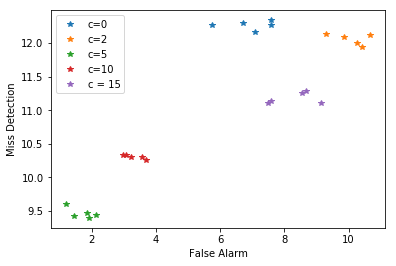}
      \caption{Performance of anomaly detection using k-NN classifier.}
          \label{fig:knn_perf}
\end{figure}

Let us now study the performance of the classifier for different types of anomalies. The false alarm and miss detection probabilities in percentages is represented in Figure \ref{fig:knn_perf}. The plot considers the anomaly detection performance for unbiased ($c=0$) and varying degrees of bias ($c>0$).

A variety of important characteristics are to be noted:
\begin{enumerate}
\item First for $c=0$, the anomalous sources behave like unbiased noisy compute nodes and thus are harder to detect, as indicated by its high miss detection and false alarm probabilities.
\item For small amounts of bias, such as $c=2$, the false alarm probability worsens with minimal to no improvement in miss detection probability. This is owing to the fact that the biased anomalies create few more invalidations among honest sources, making the classifier misclassify them as anomalies. However, the bias being small also implies that the classifier does not get much more information about the anomalies than in the unbiased case---hence the comparable miss detection probabilities.
\item Next, as the bias increases to $c=5$, we observe that the anomaly detection gets significantly easier as the bias becomes more recognizable through the invalidation profiles.
\item However, when the bias becomes significantly large, for instance $c\in\{10,15\}$, the anomalous sources are so heavily biased that they end up creating far more frequent invalidations, which make the classifier categorize honest sources as anomalies. Thus, the false alarm probability increases. The miss detection probability also grows as the anomalies are now well-hidden amongst more honest sources with similar invalidation profiles.
\end{enumerate}

\begin{figure}
	\centering
	  \includegraphics[width = 0.7\linewidth]{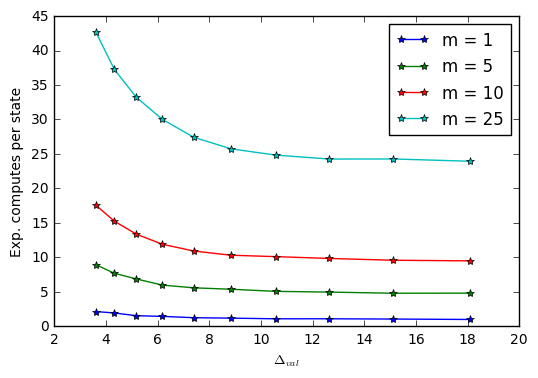}
	\caption{Computational cost}
	\label{fig:comp_cost}
\end{figure}

Separate from the task of detecting anomalous workers, it is also important to benchmark the overheads in terms of computation and communication as incurred by the validation engine. Figure \ref{fig:comp_cost} considers the average number of recomputes (or approximate computes) per simulation as a function of the validation tolerance, $\Delta_{val}$, for varying sizes of endorser sets, $m$.  With decreasing validation tolerance, $\Delta_{val}$, the computation overhead increases sharply. 

When $m$ is small, validation translates to a requirement of collision in recomputation. Since the OM computational framework incurs significant variance, the collision probability is also lower therein increasing the chance of invalidation. Thus, the scaling of computational cost for a given $\Delta_{val}$ is not linear with $m$, especially for small values of $\Delta_{val}$.

\begin{figure}
	\centering
	  \includegraphics[width = 0.7\linewidth]{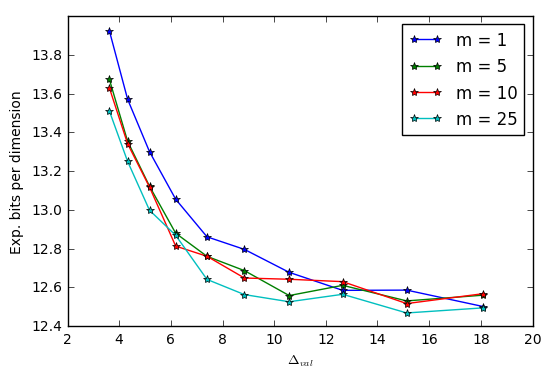}
	\caption{Communication cost}
	\label{fig:comm_cost}
\end{figure}

The compression schema enforced by the validation engine ensures a significant reduction in the average communication cost per simulation, per endorser, per dimension, as shown in Figure \ref{fig:comm_cost}. To note the reduction in cost, observe that the average number of bits per simulation, per endorser, per dimension without compression costs $32$ bits. The communication cost increases sharply with decreasing validation tolerance and decreases with the number of endorsers. This is because larger endorser sets result in fewer invalidations when comparing against more robust estimates of the mean. The reduced number of invalidations also imply reduced number of refinements and recomputations, i.e., lesser communication with each endorser.

The results discussed above were generated by the validation engine of the system whose main goal is to make a decision of either committing data to the ledger or flag the results as invalid. For invalid results, the engine  notifies the users with possible reasons for invalidations with suggestion for the simulations to be recomputed by other workers.  Users can then proceed with request for another simulation with possible updates if it was input dependent anomaly.

\section{Conclusion and Future Work}
\label{sec:conclusion}

In this paper, we have considered the problem of establishing distributed trust in the computations performed over a multi-agent system.  Using a novel combination of blockchain logging and consensus with distributed computing, we presented a validation and verification scheme for computational results. Beyond employing existing validation properties from the blockchain endorsement policy ---e.g., identity keys and metadata of the chaincode, we added the validation mechanism to extend the endorsement policy.

We have studied the cost overheads associated with the proposed system and highlighted its capacity to detect faulty or anomalous workers using the validation engine.  From our initial in-house evaluation, the creation of such trusted systems for distributed computation among `un-trusting' peers can allow improved collaboration along with efficient data, model, and result sharing that is critical to improved policy design mechanisms. Additionally, such systems can result in creating unified platforms for sharing data and facilitating scientific reproducibility and rigor in machine learning, computational science, and related fields \cite{Warden2018,LiptonS2018}.

\section{Acknowledgments}
This work was conducted under the auspices of the IBM Science for Social Good initiative. The authors thank Aleksandra Mojsilovi\'c and Aisha Walcott-Bryant for discussions and support.

\bibliographystyle{plain}
\bibliography{references}

\end{document}